\documentclass[twocolumn,english,aps,showpacs,superscriptaddress]{revtex4-2}
\usepackage[T1]{fontenc}
\usepackage{color}
\usepackage{amsmath}
\usepackage{amssymb}
\usepackage{graphicx}
\usepackage{wasysym}
\usepackage{esint}
\usepackage{comment}
\usepackage{physics}

\makeatletter
\usepackage{babel}
\usepackage{dsfont}
\usepackage[dvipsnames]{xcolor}
\usepackage[normalem]{ulem}

\usepackage{hyperref}
\def\equationautorefname~#1\null{Equation (#1)\null}

\def\maketitle{
\@author@finish
\title@column\titleblock@produce
\suppressfloats[t]}
\makeatother
\IfFileExists{lmodern.sty}{\usepackage{lmodern}}{}

\begin{document}

\title{Enhanced coupling of electron and nuclear spins by quantum tunneling resonances}

\author{Anatoli Tsinovoy}
\address{Solid State Institute, and Faculty of Physics, Technion-Israel Institute of Technology, Haifa 3200003, Israel}
\address{Rafael Ltd., Haifa 3102102, Israel}
%\email{anatolits@technion.ac.il}

\author{Or Katz}
\address{Rafael Ltd., Haifa 3102102, Israel}
\address{present address: Department of Electrical and Computer Engineering, Duke University, Durham, North Carolina 27708, USA}
\email{or.katz@duke.edu}

\author{Arie Landau}
\address{Institute of Advanced Studies in Theoretical Chemistry, Technion-Israel Institute of Technology, Haifa 3200003, Israel}
\address{Schulich Faculty of Chemistry, Technion-Israel Institute of Technology, Haifa 3200003, Israel}
%\email{alandau@technion.ac.il}

\author{Nimrod Moiseyev}
\address{Solid State Institute, and Faculty of Physics, Technion-Israel Institute of Technology, Haifa 3200003, Israel}
\address{Schulich Faculty of Chemistry, Technion-Israel Institute of Technology, Haifa 3200003, Israel}
%\email{nimrod@technion.ac.il}

\begin{abstract}

Noble-gas spins feature hours long coherence times owing to their great isolation from the environment, and find practical usage in various applications. However, this isolation leads to extremely slow preparation times, relying on weak spin transfer from an electron-spin ensemble. Here we propose a controllable mechanism to enhance this transfer rate. We analyze the spin dynamics of helium-3 atoms with hot, optically-excited potassium atoms and reveal the formation of quasi-bound states in resonant binary collisions. We find a resonant enhancement of the spin-exchange cross section by up to six orders of magnitude and two orders of magnitude enhancement for the thermally averaged, polarization rate-coefficient. We further examine the effect for various other noble gases and find that the enhancement is universal. We outline feasible conditions under which the enhancement may be experimentally observed and practically utilized.
\end{abstract}
\maketitle

% ----------------------INTRODUCTION------------------------
Spin polarized noble-gases are unique systems that can maintain their spin state for hours even at room-temperature. They have utility in various applications including precision sensing~\cite{precision_app1,walker2016spin,kominis2003romalis, sushkov2008production}, medical imaging of the brain and lungs \cite{albert1994biological,fain2010imaging,flors2017hyperpolarized,lung3,lung4}, neutron scattering experiments~\cite{gentile2017,spin_filter1}, the search for dark matter, physics beyond the standard model~\cite{newphysics1,newphysics2,newphysics3, bloch2021nasduck}, and potentially in quantum information applications including the generation of long-lived entanglement~\cite{anthony1993pl,QI_He3_a,QI_He3_b,QI_He3_c,QI_He3_d}.

The great isolation of noble-gas spins from the environment sets a trade off between their spin-polarization-rate and their spin-lifetime. The primary polarization processes for noble-gas spins rely on spin changing collisions with other atoms whose spins can be optically manipulated, such as metastable excited noble-gases~\cite{colegrove1963polarization,batz2011fundamentals,stoltz1996high,nikiel2013metastability, PhysRevA.81.033419,Lefevre1977,PhysRevLett.17.513} or alkali vapor in the ground-state~\cite{kastler1950some,brossel1952greation,happer1972optical,happer2010optically,walker1997RMP}. While both processes are practically useful, the former approach is mostly useful for helium and relies on electrical discharge which constantly generates plasma. The plasma limits both spin lifetime and the fraction of optically-accessible atoms~\cite{gentile2017}, thus narrowing the applicability and hindering miniaturization of this approach.

Collisions with alkali atoms benefit from higher possible densities and longer spin lifetimes. It can be applied to all noble gases, and miniaturized to a greater extent~\cite{walker2016spin}. Here, the polarization rate is determined by collisions of alkali and noble-gas pairs, illustrated in Fig.~\ref{fig:SEOP_illustration}. While heavy noble-gases can be polarized quickly, their spin-lifetime is considerably shorter than light noble-gases. $^3$He in particular, exhibits the longest spin-lifetime but also the weakest coupling to alkali spins, rendering its polarization-rate extremely slow. At typical conditions, $^3$He polarization takes many tens of hours, limiting its utility.

In cold atomic and molecular gases, interaction during collisions can be greatly enhanced by quantum Feshbach or tunneling resonances~\cite{boesten1997observation, chin2010feshbach, durr2005dissociation, bloch2008many, sikorsky2018phase, klein2017directly, tomza2019cold, tomza2015cold, henson2012observation, chefdeville2013observation, thomas2004imaging}. Tunneling resonances prolong the interaction time via the formation of quasi-bound states at particular values of the kinetic energy. At room-temperature and above however, the collision dynamics comprise many tens of partial waves, the atoms follow a thermal energy distribution, and measured cross-sections attain their classically predicted values~\cite{happer2010optically, walker1972spin}. It is therefore generally assumed that quantum resonances at ambient conditions would be negligible, and consequently, their potential application for noble-gas polarization has never been considered. %A single debated work exists on the subject~\cite{FranzNaNeArticle,happer1985comment,franz1985response. In their work the Ne was spin 0 and it was considered as a source of relaxation from the alkali. In the context of spin polarization no work was considered.

\begin{figure}[b]
\begin{centering}
        \includegraphics[clip,width=8.6cm]{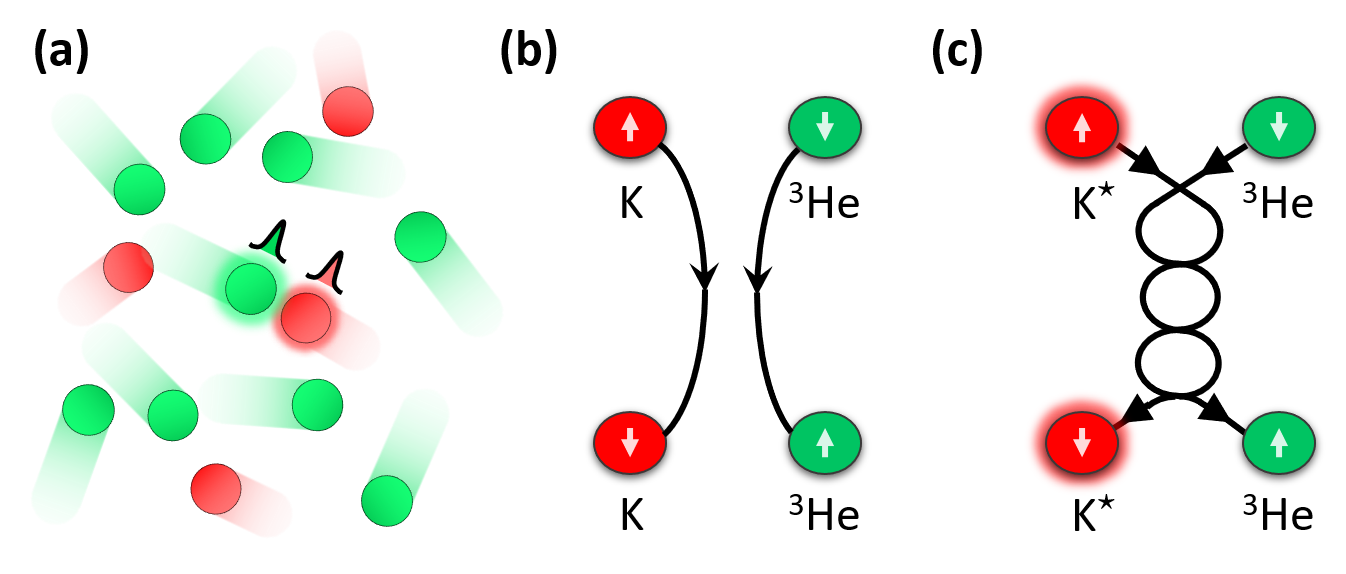}
        \par\end{centering}
\caption {\textbf{Enhancement of spin polarization-rate via quantum tunneling resonances}. (a) Alkali atoms and noble-gas atoms experience frequent spin-exchange collisions. At room temperature the quantum nature of atomic motion, such as tunneling resonances, is obscured. (b) Spin-exchange in a short binary collision between a ground-state alkali atom and a noble-gas atom. The probability of spin-exchange per collision is extremely small, resulting in slow spin-polarization rate. (c) Resonant spin-exchange between an electronically-excited alkali atom and a noble-gas atom. In resonant collisions, a quasi-bound state is formed and interaction time is increased by orders of magnitude, significantly enhancing the probability of spin-exchange.}
\label{fig:SEOP_illustration}
\end{figure} 

% ----------------------HERE WE-----------------------------

Here we propose a new mechanism to enhance the polarization-rate of noble-gas spins by resonant collisions with electronically-excited alkali atoms. We solve the quantum scattering problem of $^3$He colliding with electronically-excited potassium and reveal tunneling resonances in binary collisions~\cite{moiseyev2011non, sakurai2014modern, schultz1973resonances, zare2006resonances}. We calculate the spin-polarization rate-coefficient and find two orders-of-magnitude enhancement driven by the resonances, and up to six orders-of-magnitude enhancement of the spin-exchange cross section at specific resonance energies. We analyze the application of this mechanism to other alkali and noble-gas pairs and find universal enhancement. Finally, we outline the conditions under which the enhancement may be experimentally observed and practically utilized.

We start the analysis by solving the quantum scattering of alkali and noble-gas pairs. We then consider the energy-dependent elastic and spin-exchange cross-sections, and finally calculate the thermally-averaged rate-coefficient of the ensemble.

% For neutral gas systems, such quantum scattering is usually only observed in collisions of cold atomic beams \cite{schutte1972orbiting,grace1977observation,Toennies1974,Toennies1979}

% ------------GENERAL FORMULATION OF THE PROBLEM------------

We describe the motion of the pair during a collision using the Born-Oppenheimer approximation separating the nuclear and electronic degrees of freedom. The Hamiltonian for the relative motion of the nuclei is given by
\begin{equation}\label{eq:non_electronic_hamiltonian}
        H=-\frac{\hbar^2}{2\mu}\frac{\partial^2}{\partial R^2}+\frac{\hbar^2\mathbf{L}^2}{2\mu R^2}+V\left(R\right)+\hbar\alpha(R)\mathbf{I}\cdot\mathbf{S}
.
\end{equation}
The first two terms describe the kinetic energy where $R$ denotes the internuclear distance, $\mathbf{L}^2$ denotes the rotational angular momentum of the relative motion of the nuclei with eigenvalues $l(l+1)$, and $\mu$ denotes the reduced mass of the pair. The third term $V\left(R\right)$ describes the spin-independent potential energy curve (PEC), and the last term is the spin dependent interaction, dominated by the isotropic Fermi contact term ~\cite{walker1997RMP}. This interaction is responsible for the spin-polarization transfer from the electronic spin of the potassium atom, $\mathbf{S}$, to the nuclear spin of the helium, $\mathbf{I}$, via the hyperfine-coupling coefficient $\alpha\left(R\right)$. 

% --WE CALCULATE THE PECS AND ALPHA WITH AB-INITIO METHODS--

We calculate the \textit{ab initio} values of $V\left(R\right)$ and $\alpha\left(R\right)$ for the K-$^3$He complex as shown in Fig.~\ref{fig:PECs_vs_alphas}(a) for the $X^2\Sigma\left(\mathrm{4S}\right)$ ground-state and $^2\Sigma$ excited states.
The wave-functions of the isolated K atom and the K-$^3$He complex are constructed hierarchically. First we solve the restricted HF equations for the $($K-$^3$He$)^+$ cation (a closed shell system that serves as a reference function). We then refine the results by introducing correlations using the equation-of-motion coupled-clusters (EOM-CC) method at the singles and doubles level of theory. Finally, the valence electron is added via electron attachment  (EOM-EA-CCSD)~\cite{hirata2000high,krylov2008equation}. The calculations are preformed via the electronic-structure package Q-Chem~\cite{QCHEM4}, with $V\left(R\right)$ and the electronic wavefunction $\ket{\Psi\left(R\right)}$ as outputs, where the latter is used to calculate $\alpha(R)$ directly \cite{walker1997RMP,Bucher_2000}. % We use the latter to calculate the $\alpha(R)$ and $\alpha\left(R\right) = \expval{\Psi\left(R\right)\mid2\mu_\mathrm{I}\hat{\rho}_\mathrm{spin,He}/{3}\mid\Psi\left(R\right)}$, where $\mu_\mathrm{I}$ is the magnetic moment of the $^3$He nucleus and $\hat{\rho}_\mathrm{spin,He}$ is the spin density operator evaluated at the position of the helium nucleus. 
 Comparison with Ref.~\cite{blank_comparison} for validation and the results for the first dozen excited states are provided in Ref.~\cite{SI_ref}. %as well as the PECs and values of $\alpha\left(R\right)$ for the first dozen excited states are provided in~\cite{SI_ref}.

For collisions of helium and potassium in the ground-state, the $X^2\Sigma\left(\mathrm{4S}\right)$ potential is purely repulsive and supports no bound or quasi-bound states. In contrast, the excited-state $^2\Sigma\left(\mathrm{5S}\right)$ potential exhibits a potential well preceded by a barrier. The barrier is significant even for s-wave collisions ($l=0$), in the absence of a centrifugal potential. These wells and barriers give rise to bound states (E<0) and quasi-bound states~\cite{sakurai2014modern, moiseyev2011non} (E>0) as shown in Fig.~\ref{fig:resonances}(a). The wave-functions of both the bound and quasi-bound states (square-integrable ro-vibrational solutions), were obtained by the method of complex-scaling~\cite{moiseyev2011non} and are presented superimposed on the PEC.

\begin{figure}[t]
\begin{centering}
        \includegraphics[width=8.6 cm]{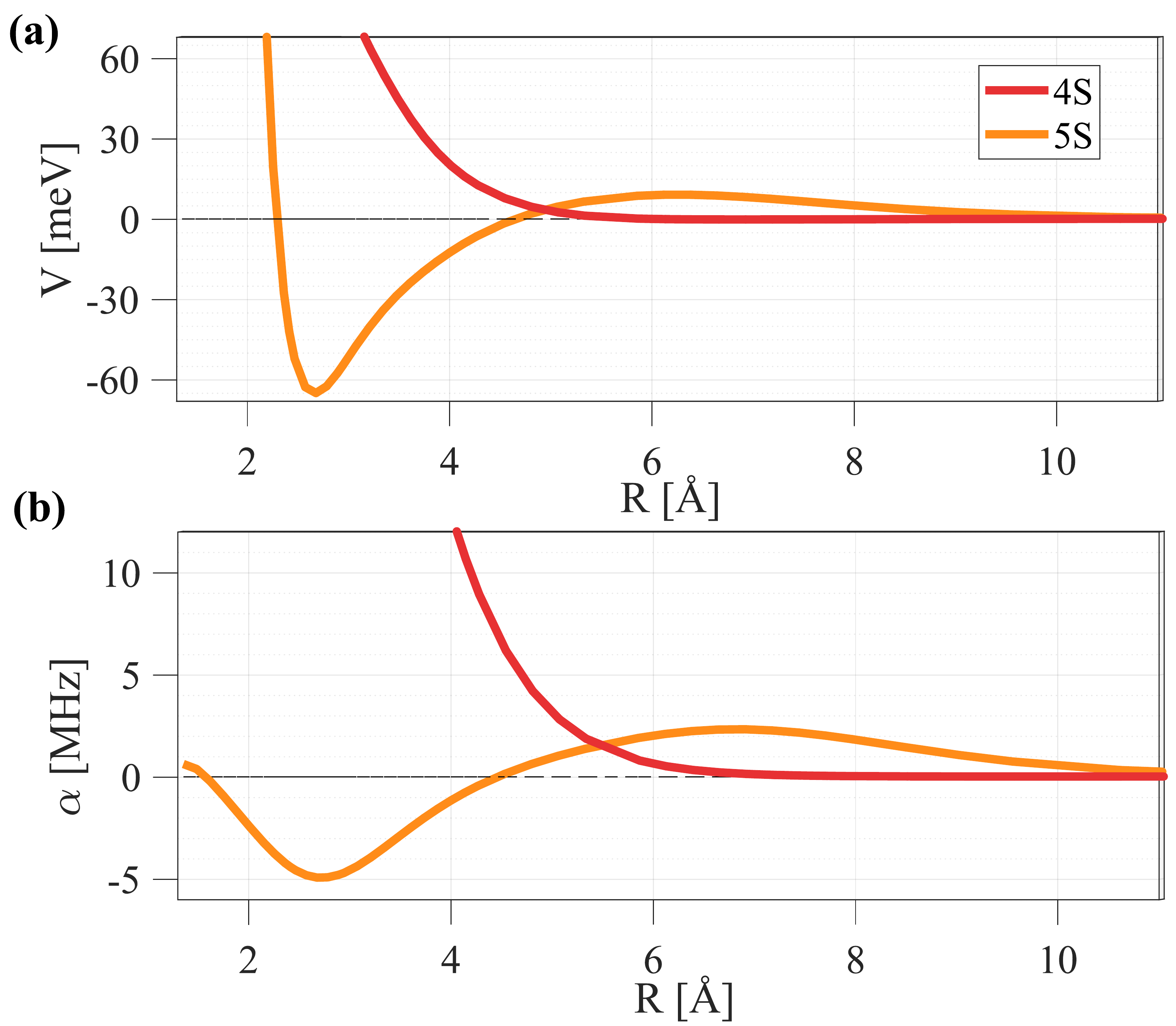}
        \par\end{centering}
         \centering{} \caption {\textbf{
                  Potential energy curves (PEC) and hyperfine-coupling coefficient for helium-3 and electronically-excited potassium.} (a) PEC of the $\mathrm{K-}^3\mathrm{He}$ complex, corresponding asymptotically to the helium atom in its ground-state and to the potassium atom in its 4S ground-state, or its 5S electronically excited states. The ground-state potential (red) is purely repulsive, whereas the excited state (orange) exhibits a potential well and a barrier. (b) Hyperfine-coupling coefficient $\alpha\left(R\right)$. The sign of $\alpha\left(R\right)$ simply indicates the precession direction of the spins.}
\label{fig:PECs_vs_alphas}
\end{figure}

\begin{figure*}[t]
\begin{centering}
        \includegraphics[ width=1\textwidth]{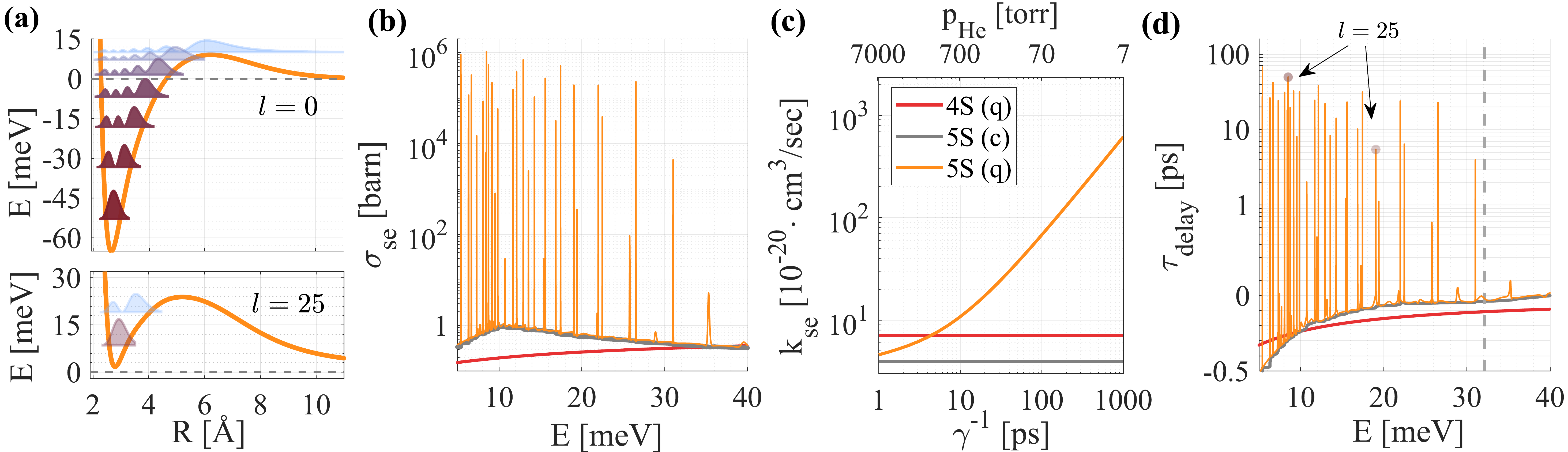}
        \par\end{centering}
         \centering{}  \caption {\textbf{Tunneling resonances in binary collisions.} (a) PEC for the 5S state with $l=0$ (top) and $l=25$ (bottom), superimposed with the wave-function of bound states (E<0) and quasi-bound states (E>0) at their resonance energies. %The latter, act as capture resonances for binary collisions of K$^\star$ with $^3$He. 
         (b) The spin-exchange cross-section $\sigma_\mathrm{se}^\mathrm{(q)}(E)$ of the 5S state (orange) is dramatically enhanced at the resonances, by up to six orders of magnitude with respect to that cross-section absent the resonances, $\sigma_\mathrm{se}^\mathrm{(c)}
         (E)$, (grey). (c) Thermally-averaged polarization rate-coefficient $k_{\mathrm{se}}^\mathrm{(q)}$ at $100\,\mathrm{^\circ C}$. $k_{\mathrm{se}}^\mathrm{(q)}$ for collisions of $^3$He with  K$^\star$ in the 5S state (orange) is enhanced by up to two orders of magnitudes by the resonances at low pressures. (d) Scattering time-delay, relative to a hard-wall potential at the origin, for a binary collision at energy $E$ and $j=0$. Sharp peak resonances signify the formation of quasi-bound states, enhancing the typical duration of semi-classical binary collisions ($0.1\,\mathrm{ps}$) by three orders of magnitude. Vertical dashed lines in (d) mark the mean thermal energy at $100\,\mathrm{^\circ C}$, and arrows exemplify the dominant resonant contribution of a specific partial wave $l=25$ to the time delay. Grey lines in (b)-(d)  present semi-classical estimations for the 5S state, which exclude the contribution of the resonances. The semi-classical limit is reached for high values of $\gamma$, where the resonances are suppressed and the collisions are entirely classical.}
\label{fig:resonances}
\end{figure*}

% -----------WE USE SPS TO SOLVE THE SCATTERING-------------

To quantify the contribution of the quasi-bound states to the polarization-rate, we solve the quantum scattering problem via the method of Siegert Pseudo States~\cite{batishchev2007siegert} which is suitable for single-channel problems. We exploit the symmetry of $\hbar\alpha\left(R\right)\mathbf{I}\cdot\mathbf{S}$ in Eq.~\eqref{eq:non_electronic_hamiltonian}, which is diagonal with respect to the joint angular momentum operator $\mathbf{J}^2\equiv\left(\mathbf{I}+\mathbf{S}\right)^2$ with eigenvalues $j\left(j+1\right)$, and solve the scattering of the singlet and triplet channels independently.

For each single channel problem, we use $N=200$ basis functions (Jacobi polynomials~\cite{SPS_onechannel}) to discretize the problem and construct a martix representation of Eq.~\eqref{eq:non_electronic_hamiltonian}. We truncate the problem at $a=40a_{0}$, explicitly approximating $V\left(R\right)\approx0$ for $R>a$, having verified convergence. Diagonalization of this matrix yields a discrete set of complex wave-numbers $k_{n,l,j}$ associated with all incoming and outgoing collision states, including the long-lived tunneling resonances. To account for shortening of the resonance-lifetimes by other processes, we introduce a relaxation rate $\gamma$ into the calculation by $\tilde{k}_{n,l,j}= \mathrm{Re}\left(k_{n,l,j}\right)+i\left[\mathrm{Im}\left(k_{n,l,j}\right)-\gamma/\left|k_{n,l,j}\right|\right]$, where $\gamma$ describes the external dissociation rate. We model $\gamma = \gamma_0 + a p$ to account for spontaneous emission at a typical rate $\gamma_0^{-1}\approx10\,\mathrm{ns}$, and for collisions with background atoms at characteristic pressure $p$ at room-temperature. While the molecular dissociation rate of stable alkali-noble gas molecules in the S manifold is about $1\,\mathrm{MHz}/\mathrm{Torr}$ ~\cite{happer1984polarization}, here we consider a more stringent rate which bounds the dissociation rate of quasi-bound molecules due to collisions with a second helium atom. As the charge density of helium is strongly localized, perturbations to the resonance states can occur only when the second helium atom overlaps with the $\mathrm{K-}^3\mathrm{He}$ wavefunction, at most 5 $\text{\AA}$ from its center, yielding $a\lesssim \sigma_\mathrm{hard-sphere}v=25\times2\pi\,\mathrm{MHz}/\mathrm{Torr}$. The partial scattering amplitudes are then given by~\cite{batishchev2007siegert}
%($\approx3\cdot10^{-11}$ cm$^3$/sec)
\begin{equation}\label{eq:S_from_kn}
    S_l^j\left(E\right)=e^{-2i\sqrt{2E}a}\prod_{n=1}^{2N+l}{\frac{\tilde{k}_{n,l,j}+\sqrt{2E}}{\tilde{k}_{n,l,j}-\sqrt{2E}}}
,
\end{equation}
and the quantum spin-exchange cross-section by the sum
\begin{equation}\label{eq:sigma_se}
    \sigma_{\mathrm{se}}^\mathrm{(q)}\left(E\right)=\frac{\pi}{8E}\sum_l{(2l+1)\left|S_l^1-S_l^0\right|^2}
    .
\end{equation}

% ----------THE SPIN-EXCHANGE CROSS-SECTION RESULT----------

We present the spin-exchange cross-section in Fig.~\ref{fig:resonances}(b) for collisions of ground-state (red) or excited-state (orange) potassium with a lifetime limit of $\gamma^{-1}=1\,\mathrm{ns}$ corresponding to $7\,\mathrm{Torr}$ of $^3$He.
The 5S state exhibits considerable increase at sharply defined peaks at specific resonant values of the kinetic energy. To highlight the role of these resonances, we first compare $\sigma_{\mathrm{se}}^\mathrm{(q)}$ with the semi-classical estimate of Ref.~\cite{walker1989estimatesPRA}, as shown in Fig.~\ref{fig:resonances}(b) (grey) and given by \begin{equation}\sigma_{\mathrm{se}}^\mathrm{(c)}
(E)=\frac{\pi}{2}\int_{0}^{\infty} bdb \Bigl|\int_{-\infty}^{\infty} \alpha\left(R\left(t\right)\right)dt \Bigr|^2.\label{eq:semi-class}\end{equation} This estimate integrates the hyperfine interaction across all possible classical collision trajectories at energy $E$.

At specific energies, the ratio $\sigma_\mathrm{se}^\mathrm{(q)} / \sigma_\mathrm{se}^\mathrm{(c)}$, for the 5S state spans up to six orders of magnitude. At room-temperature or above however, the practical polarization-rate is determined by the rate-coefficient $k_{\mathrm{se}}^\mathrm{(q)}$ which averages the spin-exchange cross-section over the Boltzmann distribution at temperature $T$. In Fig.~\ref{fig:resonances}(c), we present the polarization rate-coefficient of the ground-state (red) and excited-state (orange) at $100\,\mathrm{^\circ C}$ as a function of the inverse external dissociation rate $\gamma^{-1}$. Evidently, for long collision-lifetime-limits the excited-state polarization-rate surpasses that of the ground-state by up to two orders of magnitude. This enhancement is due to quantum resonances, as seen by comparison with the semi-classical estimate (grey) of the rate-coefficient $k_{\mathrm{se}}^\mathrm{(c)}$.

% -----------------MECHANISM VIA TIME DELAY-----------------
Before detailing the experimental proposal for measuring this enhancement and discussing its potential applications, we find it insightful to discuss its origin and its expected manifestation in other alkali and noble-gas pairs. As suggested by the semi-classical formula, Eq.~\eqref{eq:semi-class}, the cross-section is determined by the interaction strength, $\left|\alpha\left(R\right)\right|^2$, integrated over the duration of the collision. As shown in Fig.\ref{fig:PECs_vs_alphas}(b), $\left|\alpha\left(R\right)\right|$ is comparable for the ground and excited states, implying that the resonant enhancement is due to an increase in interaction time. 

We estimate the interaction time 
by calculating the temporal delay (or acceleration) of a particle with energy $E$ and angular momentum $j$ scattered by $V(R)$, relative to its free-flight time.
In Fig.~\ref{fig:resonances}(d) we present the mean time-delay $\tau_\mathrm{delay}^j\left(E\right)=\sum_l{\sigma_l^j\tau_l^j}/\sum_l{\sigma_l^j}$ for $\gamma^{-1}=1\,\mathrm{[ns]}$. $\tau_l^j=2d\delta_l^j/dE$ are the partial delays,  $\sigma_l^j\left(E\right)=(8\pi/E)\left(2l+1\right)\sin^2\delta_l^j$ are the partial elastic cross-sections and $\delta_l^j(E)=-(i/2)\log S_l^j$ are the partial scattering phase-shifts~\cite{griffiths2018introduction}. For $\mathrm{K}-^3\mathrm{He}$ in the $5\mathrm{S}$ excited-state, the time-delay features sharp peaks like those in Fig.~\ref{fig:resonances}(b), unlike the smooth ground-state response. These peaks are associated with tunneling resonances, where the mean time-delay corresponds to the lifetime of the quasi-bound state. Notably, the width of a resonance is inversely proportional to its lifetime, but its contribution to the spin-exchange cross-section scales as its lifetime squared. This allows a finite number of narrow resonances to dominate the polarization rate-coefficient. We have repeated this analysis for other estimations of the PEC~\cite{chattopadhyay2012comparative,kontar2017electronic}, and found that the enhancement of the rate coefficient remains considerable~\cite{SI_ref}.

% ---------------APPLICABILITY TO OTHER PAIRS---------------

We expect polarization enhancement via quasi-bound states to be dominant for other pairs of noble-gas and optically-excited alkali atoms. Quasi-bound states originate from wells and barriers in the shape of the PEC, which appear in various alkali noble-gas pairs~\cite{blank_comparison, masnou1982model, tam1975strong}, and are correlated with the shape of the electron density as shown for $\mathrm{LiHe}$ in Ref.~\cite{yiannopoulou1999undulations}. For most resonant collisions, the interaction time is saturated by $\gamma^{-1}$, and therefore each resonance contributes similarly to $k_{\mathrm{se}}^\mathrm{(q)}$. The enhancement is thus proportional to the total number of resonances, $N_\mathrm{res}$ which predominantly depends on $\mu$. % Heavier pairs have deeper wells, larger $\alpha$, and higher reduced mass $\mu$.
By scaling $\mu$ in Eq.~\eqref{eq:non_electronic_hamiltonian} for the $\mathrm{K-}^3\mathrm{He}$ potential we find that $N_\mathrm{res}\propto\mu$ as presented in Fig.~\ref{fig:KAr_fig}(a). We verify this estimate by solving the scattering of electronically-excited $\mathrm{K-}^{37}\mathrm{Ar}$ pairs, using an ab-initio 5S potential~\cite{zohar_pc}.  %The strong agreement implies that the specific details of the potential are of little significance. 
In Fig~\ref{fig:KAr_fig}(b) we show the increase in the number of resonances for $\mathrm{K-}^{37}\mathrm{Ar}$ as expressed in the mean time-delay. We characterize the enhancement by the resonances over binary polarization rate in Fig.~\ref{fig:KAr_fig}(c) using the ratio $k_\mathrm{se}^\mathrm{(q)} / k_\mathrm{se}^\mathrm{(c)}$ which weakly depends on $\alpha\left(R\right)$ and on the specific colliding pair.

\begin{figure}[t]
\begin{centering}
        \includegraphics[ width=1\columnwidth]{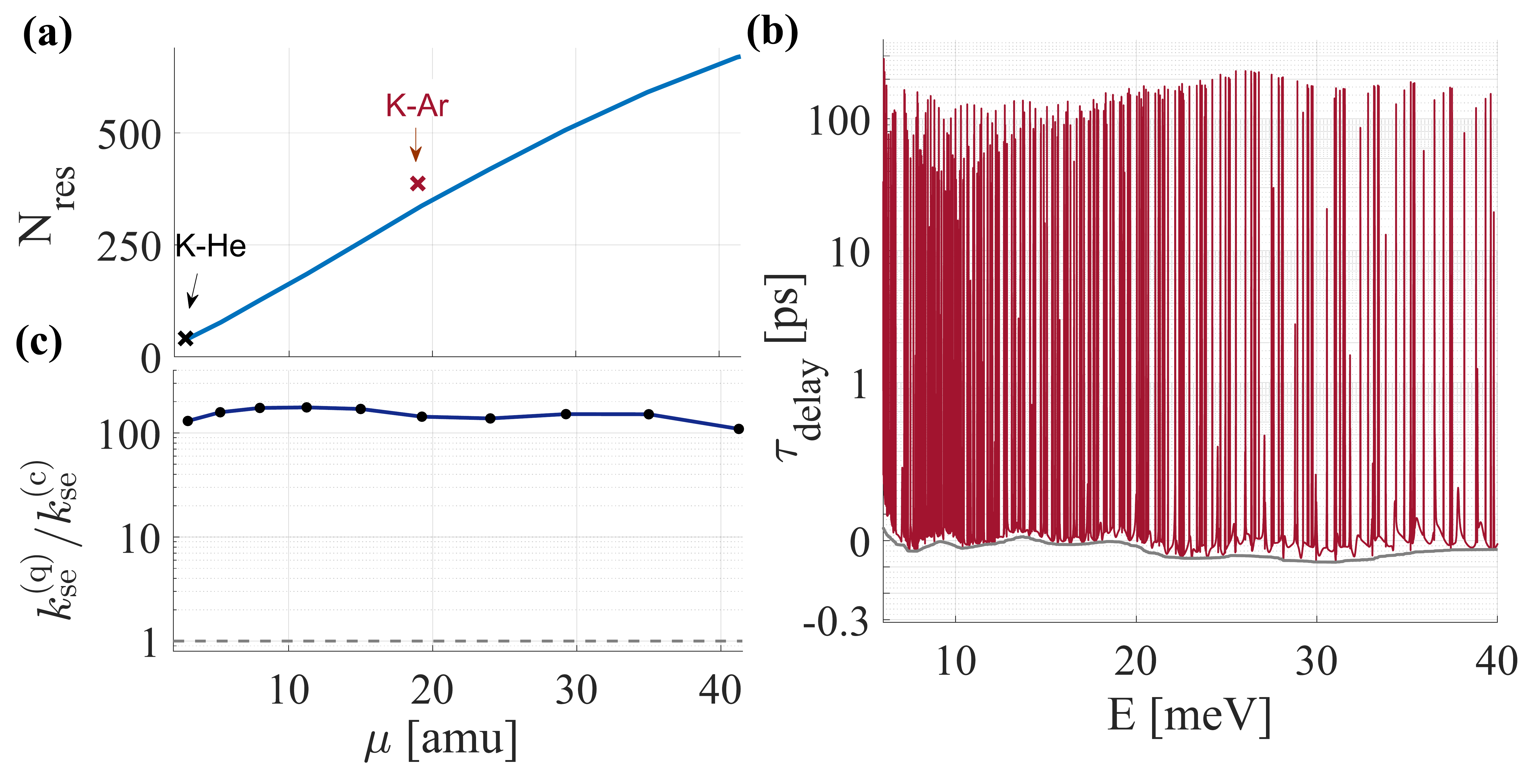}
        \par\end{centering}
         \centering{}  \caption {\textbf{Resonant spin-polarization transfer for different electronically-excited alkali-noble-gas pairs.} (a) The number of tunneling resonances $N_{\mathrm{res}}$ (blue line) increases linearly with $\mu$, derived by solving the $\mathrm{K-}^3\mathrm{He}$ scattering in Eq.~\eqref{eq:non_electronic_hamiltonian} scaled by $\mu$. Crosses mark calculations for unscaled $\mathrm{K-}^3\mathrm{He}$ (black) and the true 5S PEC of $\mathrm{K-}^{37}\mathrm{Ar}$ (brown).
         (b) Scattering time-delay, for the heavier $\mathrm{K-}^{37}\mathrm{Ar}$ pair.
         The number of resonances is dramatically increased compared with Fig.~\ref{fig:resonances}(d).
         (c) The enhancement in the polarization rate-coefficient $k_\mathrm{se}^\mathrm{(q)} / k_\mathrm{se}^\mathrm{(c)}$ for the 5S potential by formation of tunneling resonances over binary collisions which we estimate to be constant and independent of $\mu$ and $\alpha$.}
\label{fig:KAr_fig}
\end{figure}

% ------------------EXPERIMENTAL PROPOSAL-------------------

Resonant spin-exchange optical-pumping of noble-gas can be realized with various alkali and noble-gas mixtures, within a large range of experimental parameters. Here we present an exemplary configuration using a mixture of $\mathrm{K-}^{3}\mathrm{He}$ in a chip-scale cubic cell of length $2\,\textrm{mm}$ and $3\,\mathrm{Torr}$ of helium. The resonant polarization transfer relies on optical pumping of potassium spins in the 5S state followed by random collisions with helium gas. The $5\mathrm{S}$ state can efficiently be excited via a two-photon ladder scheme $4\text{S}\rightarrow4\text{P}_{3/2}\rightarrow5\text{S}$ using $766\,\mathrm{nm}$ and $1.25\,\mathrm{\mu m}$ light for the first and second transitions, respectively. The 4P level is pressure-broadened, with homogeneous optical line-width of about $\Gamma\approx50\,\mathrm{MHz}$ ~\cite{happer2010optically} and inhomogeneous Doppler broadening of $1\,\mathrm{GHz}$. The dominant decay rate of the 5S state, $\gamma_{0}=3.8\,\mathrm{MHz}$ is radiative, as the coupling of S shells to orbital angular momentum is weak, suppressing destruction by spin-rotation during collisions.

$500\,\textrm{mW}$ for each beam covering the cell yields Rabi frequencies exceeding $300\,\textrm{MHz}$ and $100\,\textrm{MHz}$. A $3\,\mathrm{GHz}$ one-photon detuning of the beams renders the transition 4S$\rightarrow$5S through P level virtual, suppressing spin-relaxation by collisions in the P state, yielding a Raman rate $\Omega_R>10\,\mathrm{MHz}$. Pulsed operation potentially enables near unity population as $p_{5\mathrm{S}}\approx\Omega_R^2/(\Omega_R^2+\gamma_{0}^2/2)$ and about half that value for CW operation. 
The spin state of the 5S can be defined via controlling the polarizations of the beams and pumping of the 4S spin. For example, setting $766\,\mathrm{nm}$ light polarization circular, pumps the 4S spin and overcomes its $100\,\mathrm{kHz}$ depolarization rate by collisions with the walls, whereas the $1.25\,\mathrm{\mu m}$ light can be linearly polarized. If necessary, additional on-resonance pulses can maintain the $4\mathrm{S}$ spin-polarized and enable well-defined excitation channels.
For the low pressure configuration, the stringent estimation of $\gamma^{-1}=1\,\mathrm{ns}$ corresponds to a 100-fold enhancement of the polarization rate-coefficient via resonant collisions (cf.~Fig.~\ref{fig:resonances}c).

% ----------------------------SUMMARY-----------------------

% TODO - Summary as is?
In summary, we analyzed the spin-polarization transfer in collisions of optically-excited alkali and noble-gas atoms, using ab-initio calculations of K-He and K-Ar pairs. We revealed the formation of quasi-bound states, manifested as sharp resonances in the scattering time-delay and spin-exchange cross-section. The resonances are expected to enhance the polarization transfer-rate of noble gases by two orders of magnitude for a thermal ensemble at ambient conditions and up to six orders of magnitude at the resonance energies, and be significant for different optically-excited alkali and noble-gas pairs.

% --------------------------APPLICATIONS--------------------
Various applications using spin-polarized gases can benefit from optically-controlled enhancement of the polarization rate. Here we consider several potential avenues. Precision NMR sensors and co-magnetometers use mixtures of noble-gas and alkali spins. The former sense external fields, the latter serve as an embedded optical magnetometer~\cite{8358161, fang2013atomic, WALKER2016373, PhysRevLett.120.033401, PhysRevD.100.062004, PhysRevLett.121.031101}. In miniaturized sensors, with significant alkali polarization loss to cell walls, initialization time and sensitivity can greatly benefit from enhanced polarization rates for all noble-gases. This is much like the case of xenon which can be quickly polarized and thus suitable for miniaturization~\cite{jimenez2014optical}. Notably, the proposed mechanism is optically-controlled, enhancing the polarization rate on-demand within the standard operation of these sensors.

MRI of human air spaces with record resolution, and preparation of neutron spin filters and targets use large volumes of polarized gas at atmospheric pressure or above. The method of Metastability-Exchange Optical Pumping (MEOP) enables rapid polarization of helium nuclei at low gas pressures. Subsequent compression then brings the polarized gas to a higher target pressure~\cite{gentile2017}. Whereas MEOP is exclusively limited to helium, the proposed technique enable quick low-pressure polarization of other noble-gas atoms, which may be more available or more appropriate for specific applications.

Several quantum information applications, such as optical quantum memories~\cite{QI_He3_d,QI_He3_c}, generation of spin entanglement~\cite{reinaudi2007squeezing,QI_He3_b, PhysRevLett.127.013601} and nonclassical coupling to opto-mechanical systems like gravitational-wave detectors~\cite{moller2017quantum,thomas2020entanglement,khalili2018overcoming} can significantly benefit from the long spin-lifetime of noble-gases. These applications require a bi-directional interface between spins and light, and overcoming classical noise to reach the standard quantum limit~\cite{shaham2021strong, Katzeabe9164}. An efficient interface requires several ingredients including high spin-polarization, increased number densities and strong optical interaction. The classical limiting noise typically scales with the number of noble-gas atoms in the cell. Therefore, quantum applications are likely to be first realized at low pressures and at small volumes, conditions under which the proposed mechanism is most beneficial.

Finally, the resonant enhancement at particular kinetic energies is several orders of magnitude greater than the thermally-averaged one. Cryogenic operation and usage of velocity selective atomic beams might exploit that enhancement even further.%, enabling direct observation of individual tunneling resonances.

\begin{acknowledgments}

We thank Or Peleg, Roy Shaham, Yoav Sagi and Edvardas Narevicius for fruitful discussions. We also thank Zohar Amitay for making available their results for the potassium-argon PEC. This research was supported by the Israel Science Foundation (Grant No. 1661/19).

\end{acknowledgments}
\nocite{partridge2001potential,K_basis_set,EMSL,davidson2012reduced,Kutzelnigg1988,schaefer1989frequency,babcock2005he,babcock2003hybrid,babcock2005spin,walker2010method,singh2015development,spin_densities,singh2015development}

\clearpage
\newpage
\makeatletter
%%%%%%%%%%%%%%%%%%%%%%%%%%%%%% User specified LaTeX commands.

\renewcommand{\theequation}{S\arabic{equation}}
\renewcommand{\thefigure}{S\arabic{figure}}
\renewcommand{\thetable}{S-\Roman{table}}

\renewcommand{\citenumfont}[1]{S#1} 

\setcounter{figure}{0}    
\setcounter{page}{1}

\makeatother

\title{Supplementary Information for:\\
Enhanced coupling of electron and nuclear spins by quantum tunneling resonances}

\maketitle
\onecolumngrid

% ----------------------------------------------------------------------------------------------------------------------------------------------------------------------------------
\section*{S1. Ab Initio Calculation for the K-$^3$He Complex}

In this section, we provide additional details about the \textit{ab initio} calculation of the potentials and hyperfine coupling coefficients using the Q-chem package~\cite{QCHEM4}, as described in the main text.

The basic scheme of the calculation for the K-$^3$He complex follows the following stages: first, we choose a basis set and calculate the corresponding atomic orbitals from its specification, supplying them as *.in files to Q-Chem. We used the aug-pc-3 basis set \cite{partridge2001potential,K_basis_set}, taken from the EMSL basis set exchange~\cite{EMSL}. This basis provided good convergence compared to experimental values as presented in Table~\ref{tab:convergence_of_augpc3}. Second, we run a restricted HF calculation for the closed-shell and paired spin $\left(\mathrm{K-}^3\mathrm{He}\right)^+$ complex, generating a set of molecular orbitals used to construct the total electronic wavefunction, $\left\{\phi_p\left(\mathbf{r}\tau\right)\right\}$, which are functions of a single set of electron position and spin coordinates $\mathbf{r}\tau$. For a restricted HF calculation of a spin-paired system, the resulting orbitals are equal for $\tau=\alpha$ and $\tau=\beta$. These orbitals are a linear combination of atomic orbitals, and their coefficients are listed in the *.fchk files generated by Q-Chem. Third, a coupled-cluster (CC) calculation introduces electronic correlations to the $\left(\mathrm{K-}^3\mathrm{He}\right)^+$ complex, generating a correlated reference state. Finally, a single spin-up electron is attached to the reference state using electron attachment (EA) by the method of EA-EOM-CC~\cite{hirata2000high,krylov2008equation}. All CC calculations are performed at the singles + doubles (SD) level of theory.

Within the confines of the Born-Oppenhemier approximation, we repeat the calculation for several values of $R$, the intermolecular distance between the $^3$He and K nuclei. We retrieve the energies of the complex as a direct output of the computation, stored in the *.out files.

In atomic units, the hyperfine couping constant is $\alpha\left(R\right)=2\mu_\mathrm{I}\rho_\mathrm{spin}\left(R\right)/{3}$, where $\mu_\mathrm{I}$ is the magnetic moment of the $^3$He nucleus, and $\rho_\mathrm{spin}\left(R\right)$ is the total spin density function of the complex for intermolecular distance $R$, calculated at the position of the helium nucleus~\cite{Bucher_2000}.

We extract the total electronic spin density of the complex by following the standard procedure~\cite{spin_densities},  $\rho_{\mathrm{spin}}\left(R\right)=\sum_{p,q}\phi^*_p(R)\phi_q(R)\left(D_{pq}^{\alpha}-D_{pq}^{\beta}\right)$ where the molecular orbitals are evaluated at the position of the $^3$He nucleus, and $D_{pq}^\tau$ are the expectation values over the second-quantization operators $a_{p\tau}^\dagger a_{q\tau}$ for $\tau\in\{\alpha,\beta\}$.
The values of $D_{pq}^{\alpha}$ and $D_{pq}^{\beta}$ are directly output by Q-Chem and stored in the FILE\#.6 temporary files (after enabling the flag for saving temporary files).

Note that our calculation does not include the spin orbit coupling, which is not important for light nuclei beyond the asymptotic fine structure splitting.

%We calculations for the isolated potassium and helium atoms, as well as for the and $\mathrm{K-}^3\mathrm{He}$ complex.
%Our calculation does not include the spin orbit coupling, which are usually not important for light nuclei beyond the asymptotic fine structure splitting.

\begin{table}[ht]
\centering
\begin{tabular}{|l|l|l|l|}
\hline
State & E$_{\mathrm{exp}}$ [eV] &  E$_{\mathrm{aug-pc-3}}$ [eV] & $\delta\equiv\frac{E_{\mathrm{aug-pc-3}}-E_{\mathrm{exp}}}{|E_{\mathrm{exp}}|}$ \\ \hline \hline
4s & -4.3407 & -4.3027 & 0.7\% \\ \hline
4p & -2.7307 & -2.6975 & 1.2\% \\ \hline
5s & -1.7337 & -1.7236 & 0.6\% \\ \hline
3d & -1.6707 & -1.6386 & 1.9\% \\ \hline
5p & -1.2781 & -1.0339 & 19\% \\ \hline
4d & -0.9439 & -0.9132 & 3.2\% \\ \hline
\end{tabular}
\caption{Convergence of energy for the aug-pc-3 basis set. aug-pc-3 is satisfactory for all but the 5p states.\label{tab:convergence_of_augpc3}
}
\end{table}
\clearpage

\section*{S2. Potential Energy Curves and Values of $\alpha$}

\begin{table*}[ht!]
\resizebox{0.75\textwidth}{!}{%
\begin{tabular}{|l|l|l|l|l|l|l|l|l|l|l|}
\hline
\textbf{R} &
  \textbf{4S : $^2\Sigma$} &
  \textbf{4P : $^2\Sigma$} &
  \textbf{4P : $^2\Pi$} &
  \textbf{5S : $^2\Sigma$} &
  \textbf{3D : $^2\Sigma$} &
  \textbf{3D : $^2\Pi$} &
  \textbf{3D : $^2\Delta$} &
  \textbf{5P : $^2\Sigma$} &
  \textbf{5P : $^2\Pi$} &
  \textbf{Cation} \\ \hline
\textbf{2.5}  & 6491.3     & 6284.2      & 6243.1      & 6284.4      & 6473.3      & 6176.4     & 6278.8      & 6264.5     & 6239.6      & 6291.9     \\ \hline
\textbf{2.75} & 3975.2     & 3777.2      & 3676        & 3720        & 3899.2      & 3617.8     & 3711.5      & 3727.7     & 3673.7      & 3718.1     \\ \hline
\textbf{3}    & 2406.1     & 2240.9      & 2086.4      & 2128.8      & 2305.2      & 2034.3     & 2121.6      & 2159       & 2085.8      & 2123.7     \\ \hline
\textbf{3.25} & 1453       & 1330.6      & 1131        & 1169.1      & 1346.9      & 1083.3     & 1165.1      & 1215.7     & 1131.9      & 1164.1     \\ \hline
\textbf{3.5}  & 885.61     & 809.15      & 572.52      & 604.88      & 787.17      & 528.25     & 604.93      & 662.58     & 575.03      & 601.83     \\ \hline
\textbf{3.75} & 552.21     & 520.94      & 255.24      & 281.2       & 469.9       & 213.79     & 285.37      & 345.93     & 259.1       & 281.04     \\ \hline
\textbf{4}    & 357.71     & 368.42      & 81.275      & 100.9       & 297.2       & 42.318     & 108.82      & 169.45     & 86.262      & 103.85     \\ \hline
\textbf{4.2}  & 262.49     & 303.41      & 4.3829      & 19.298      & 222.12      & -32.7      & 29.755      & 88.998     & 10.071      & 24.565     \\ \hline
\textbf{4.4}  & 200.58     & 268.21      & -38.308     & -27.615     & 181.85      & -73.566    & -15.112     & 41.764     & -32.093     & -20.328    \\ \hline
\textbf{4.5}  & 178.2      & 257.77      & -50.919     & -42.124     & 170.65      & -85.277    & -28.793     & 26.579     & -44.504     & -33.97     \\ \hline
\textbf{4.6}  & 160.02     & 250.57      & -59.301     & -52.251     & 163.78      & -92.764    & -38.223     & 15.445     & -52.725     & -43.334    \\ \hline
\textbf{4.8}  & 132.96     & 242.52      & -66.908     & -62.885     & 159.43      & -98.602    & -47.85      & 1.8145     & -60.126     & -52.769    \\ \hline
\textbf{5}    & 114.17     & 238.85      & -66.692     & -65.088     & 163.1       & -96.646    & -49.534     & -4.6921    & -59.843     & -54.209    \\ \hline
\textbf{5.2}  & 100.29     & 236.32      & -62.321     & -62.558     & 170.91      & -90.57     & -46.928     & -7.7876    & -55.528     & -51.333    \\ \hline
\textbf{5.4}  & 89.194     & 233.04      & -56.113     & -57.675     & 180.39      & -82.698    & -42.352     & -9.8728    & -49.482     & -46.477    \\ \hline
\textbf{5.5}  & 84.277     & 230.8       & -52.78      & -54.83      & 185.21      & -78.55     & -39.784     & -11.02     & -46.264     & -43.771    \\ \hline
\textbf{5.75} & 73.15      & 223.18      & -44.584     & -47.399     & 196.51      & -68.361    & -33.35      & -14.998    & -38.436     & -37.005    \\ \hline
\textbf{6}    & 62.905     & 212.49      & -37.418     & -40.461     & 205.49      & -59.305    & -27.757     & -21.307    & -31.734     & -31.094    \\ \hline
\textbf{6.25} & 53.474     & 199.47      & -31.397     & -34.244     & 211.59      & -51.466    & -23.116     & -29.558    & -26.229     & -26.166    \\ \hline
\textbf{6.5}  & 44.825     & 184.85      & -26.494     & -28.828     & 214.41      & -44.842    & -19.424     & -38.777    & -21.875     & -22.213    \\ \hline
\textbf{6.75} & 37.075     & 169.45      & -22.495     & -24.088     & 213.93      & -39.225    & -16.479     & -47.487    & -18.436     & -19.032    \\ \hline
\textbf{7}    & 30.303     & 153.93      & -19.169     & -19.863     & 210.36      & -34.385    & -14.066     & -54.325    & -15.661     & -16.405    \\ \hline
\textbf{7.25} & 24.532     & 138.8       & -16.341     & -16.04      & 204.13      & -30.152    & -12.024     & -58.414    & -13.363     & -14.17     \\ \hline
\textbf{7.5}  & 19.711     & 124.4       & -13.901     & -12.555     & 195.8       & -26.414    & -10.234     & -59.499    & -11.423     & -12.227    \\ \hline
\textbf{7.75} & 15.73      & 110.9       & -11.795     & -9.3907     & 185.98      & -23.114    & -8.7237     & -57.808    & -9.7771     & -10.533    \\ \hline
\textbf{8}    & 12.467     & 98.402      & -9.9855     & -6.5422     & 175.23      & -20.212    & -7.3917     & -53.816    & -8.3887     & -9.0647    \\ \hline
\textbf{8.5}  & 7.6219     & 76.465      & -7.1675     & -1.7862     & 152.73      & -15.485    & -5.2918     & -41.104    & -6.2842     & -6.755     \\ \hline
\textbf{9}    & 4.4091     & 58.463      & -5.2246     & 1.8202      & 130.78      & -11.948    & -3.9386     & -25.168    & -4.882      & -5.14      \\ \hline
\textbf{9.5}  & 2.3443     & 44.053      & -3.8711     & 4.4948      & 110.64      & -9.2764    & -2.9663     & -8.6715    & -3.9165     & -3.9974    \\ \hline
\textbf{10}   & 1.0868     & 32.77       & -2.8918     & 6.4382      & 92.787      & -7.2124    & -2.2558     & 6.5843     & -3.193      & -3.1467    \\ \hline
\textbf{11}   & -0.0043538 & 17.527      & -1.6076     & 8.617       & 63.864      & -4.3296    & -1.2955     & 28.775     & -2.1331     & -1.972     \\ \hline
\textbf{11.5} & -0.18857   & 12.604      & -1.2025     & 9.0026      & 52.423      & -3.3497    & -0.98424    & 34.678     & -1.7467     & -1.5777    \\ \hline
\textbf{12}   & -0.27103   & 8.9444      & -0.91376    & 9.0086      & 42.738      & -2.6009    & -0.75757    & 37.339     & -1.4417     & -1.2827    \\ \hline
\textbf{12.5} & -0.30205   & 6.2355      & -0.71239    & 8.7085      & 34.625      & -2.0346    & -0.59783    & 37.339     & -1.2033     & -1.0653    \\ \hline
\textbf{13}   & -0.30912   & 4.2295      & -0.57498    & 8.177       & 27.895      & -1.6087    & -0.4977     & 35.282     & -1.0172     & -0.90505   \\ \hline
\textbf{13.5} & -0.29851   & 2.7429      & -0.47049    & 7.4921      & 22.361      & -1.2838    & -0.32926    & 31.713     & -0.86342    & -0.78042   \\ \hline
\textbf{14}   & -0.28681   & 1.634       & -0.40572    & 6.7133      & 17.828      & -1.0384    & -0.36382    & 27.129     & -0.7426     & -0.68546   \\ \hline
\textbf{14.5} & -0.26558   & 0.8158      & -0.34885    & 5.9041      & 14.144      & -0.84491   & -0.3181     & 22.021     & -0.63593    & -0.60437   \\ \hline
\textbf{15}   & -0.23919   & 0.22449     & -0.29824    & 5.1073      & 11.172      & -0.65797   & -0.26123    & 16.833     & -0.54069    & -0.53307   \\ \hline
\textbf{16}   & -0.18422   & -0.45933    & -0.21851    & 3.661       & 6.9144      & -0.46014   & -0.20572    & 7.608      & -0.38368    & -0.41225   \\ \hline
\textbf{17}   & -0.14041   & -0.68056    & -0.16218    & 2.4923      & 4.2844      & -0.31484   & -0.15565    & 1.1617     & -0.27157    & -0.32382   \\ \hline
\textbf{18}   & -0.10422   & -0.63103    & -0.11919    & 1.6262      & 2.6822      & -0.21851   & -0.11619    & -2.3345    & -0.19021    & -0.25388   \\ \hline
\textbf{20}   & -0.062858  & -0.31348    & -0.071022   & 0.61144     & 0.9426      & -0.11919   & -0.070478   & -3.5889    & -0.10041    & -0.1649    \\ \hline
\textbf{22}   & -0.039457  & -0.11184    & -0.045171   & 0.18966     & 0.0849      & -0.071838  & -0.044082   & -2.2324    & -0.05796    & -0.11184   \\ \hline
\textbf{24}   & -0.021497  & -0.041089   & -0.025307   & 0.041089    & -0.0092519  & -0.04245   & -0.0087076  & -1.0183    & -0.032382   & -0.074831  \\ \hline
\textbf{34}   & 0.0013606  & -0.0013606  & -0.0013606  & -0.0051702  & 0.0021769   & -0.0065307 & 0.0027211   & -0.0048981 & -0.0021769  & -0.014694  \\ \hline
\textbf{36}   & 0.0013606  & -0.00054423 & -0.0013606  & -0.0035375  & -0.0019048  & -0.0059865 & -0.0016327  & -0.002449  & -0.0016327  & -0.011701  \\ \hline
\textbf{38}   & 0.0016327  & 0           & -0.00081634 & -0.002449   & -0.0010885  & -0.0054423 & -0.00054423 & -0.0010885 & -0.0010885  & -0.0092519 \\ \hline
\textbf{40}   & 0.0019048  & 0.00054423  & -0.00054423 & -0.0016327  & -0.00081634 & -0.0048981 & -0.00054423 & 0          & -0.00054423 & -0.0073471 \\ \hline
\textbf{44}   & 0.0040817  & 0.0029933   & 0.0016327   & 0.0010885   & 0.0010885   & -0.0027211 & 0.0013606   & 0.0027211  & 0.0016327   & -0.0032654 \\ \hline
\textbf{46}   & 0.0043538  & 0.0032654   & 0.0019048   & 0.0016327   & 0.0013606   & -0.0021769 & 0.0019048   & 0.0032654  & 0.0021769   & -0.0019048 \\ \hline
\textbf{48}   & 0.0019048  & 0.00081634  & -0.00054423 & -0.0010885  & -0.0013606  & -0.0048981 & -0.00081634 & 0.00081634 & -0.00054423 & -0.0040817 \\ \hline
\textbf{50}   & 0.0019048  & 0.00081634  & -0.00054423 & -0.00081634 & -0.0013606  & -0.0048981 & -0.00081634 & 0.00081634 & -0.00054423 & -0.0035375 \\ \hline
$\mathbf{\infty}$      & 0          & 1605.2      & 1605.2      & 2579.1      & 2664.2      & 2664.2     & 2664.2      & 3268.9     & 3268.9      & 4302.8     \\ \hline
\end{tabular}%
}
\caption{Tabular values of the potentials for the 13 lowest lying states of the K-He complex. Values of R (in bohr) and V (in meV) for all calculated states. Values of V are listed relative to their asymptotic values and values at $\infty$ are relative to the energy of $\mathrm{K-}^3\mathrm{He}$ ground state.}
\label{tab:V_table}
\end{table*}

\begin{figure*}[t]
\centering
\includegraphics[width=0.8\textwidth]{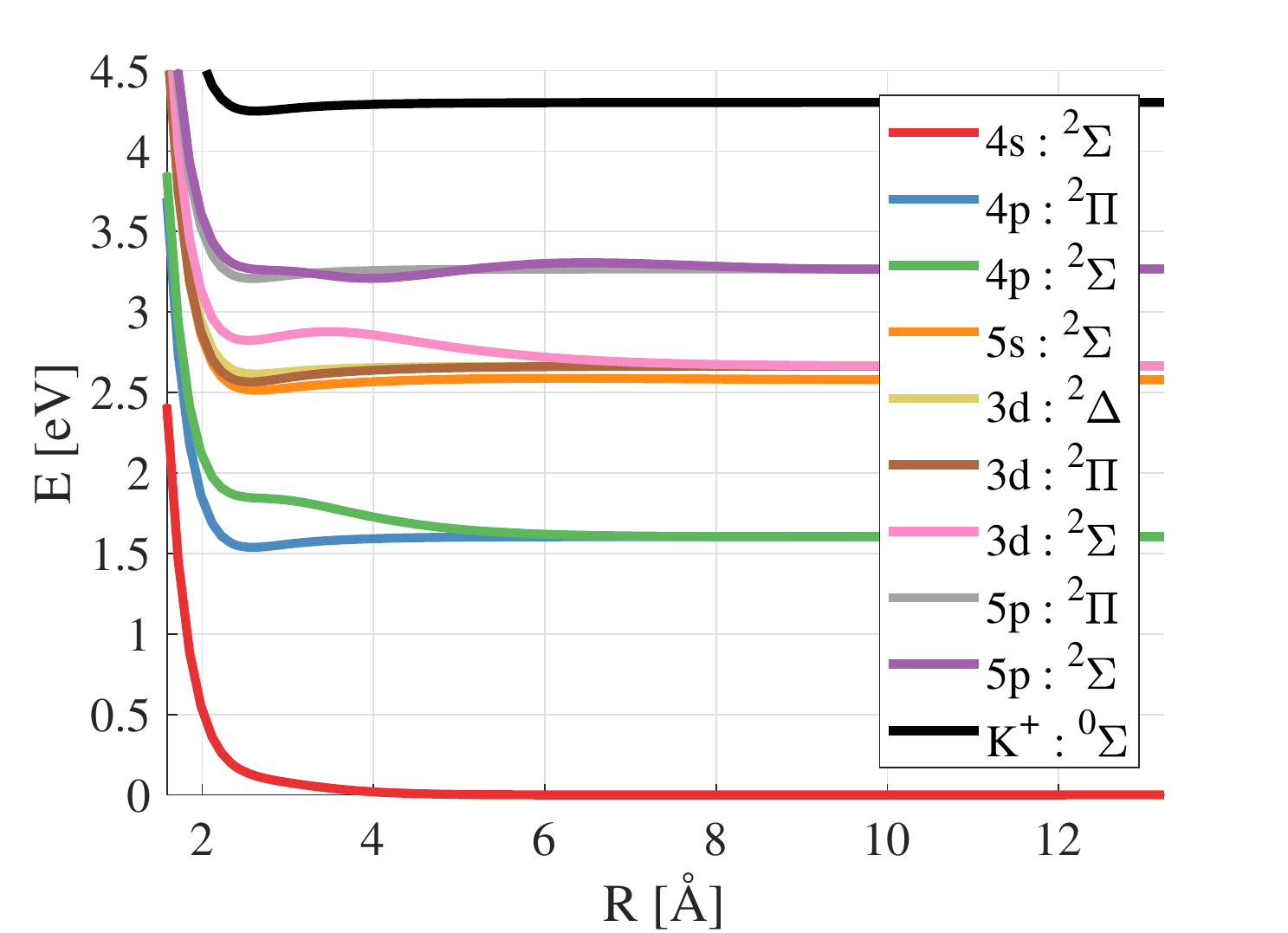}
\caption{The PECs as a function of R, the internuclear separation for all calculated states. The ground-state is predominantly repulsive, whereas the cation state exhibits a Coulombic potential well. The excited states all exhibit some structure in the form of potential wells and barriers. The wells are similar to that of the cation state (shown in main text). The longer range behavior of the $^2\Sigma$ states is repulsive, but at closer internuclear distances, there is a significant decrease in the potential energy. This forms a barrier, and for $^2\Sigma\left(\mathrm{5S}\right)$ and $^2\Sigma\left(\mathrm{5P}\right)$ a well forms. As explored in detail in the main text - this leads to the existence of shape-type resonances and truly bound states.
}
\label{fig:all_PECs_SI}
\end{figure*}

In table~\ref{tab:V_table} we list the calculated potentials of the 13 lowest lying states of the K-He complex, plotted in Fig.~\ref{fig:all_PECs_SI}.

\clearpage

\begin{figure*}[t]
\centering
\includegraphics[width=0.8\textwidth]{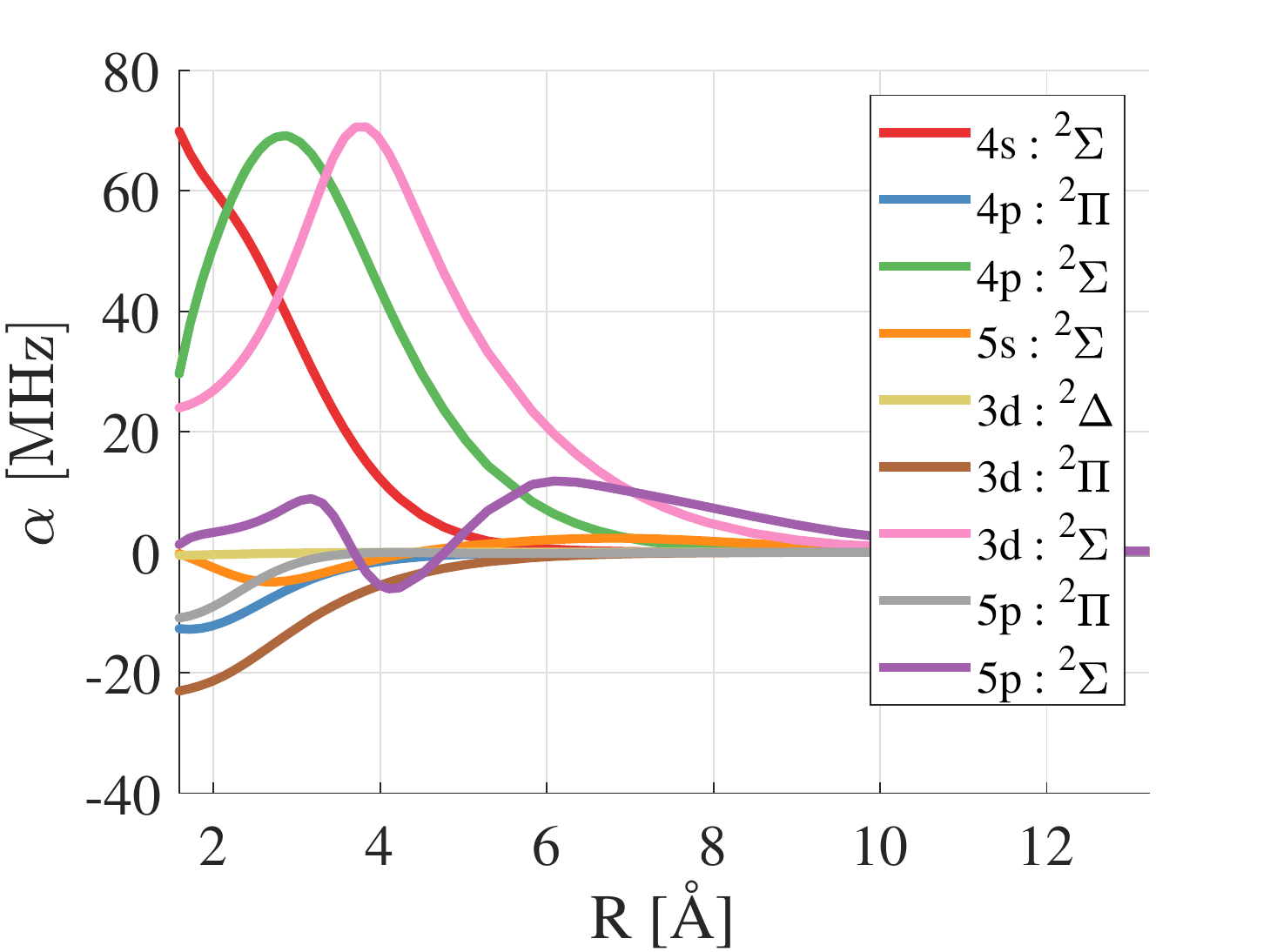}
\caption{The values of $\alpha\left(R\right)$, at internuclear separation $R$, for all calculated states.
}
\label{fig:all_rhos}
\end{figure*}

\begin{table*}[ht!]
\resizebox{0.75\textwidth}{!}{%
\begin{tabular}{|l|l|l|l|l|l|l|l|l|l|}
\hline
\textbf{R} &
  \textbf{4S : $^2\Sigma$} &
  \textbf{4P : $^2\Sigma$} &
  \textbf{4P : $^2\Pi$} &
  \textbf{5S : $^2\Sigma$} &
  \textbf{3D : $^2\Sigma$} &
  \textbf{3D : $^2\Pi$} &
  \textbf{3D : $^2\Delta$} &
  \textbf{5P : $^2\Sigma$} &
  \textbf{5P : $^2\Pi$} \\ \hline
\textbf{2.5}  & 79.391      & 2.526       & -11.664     & 0.62329     & 23.948      & -23.522     & -0.52603    & -6.5153     & -10.831     \\ \hline
\textbf{2.75} & 74.623      & 18.313      & -12.293     & 0.37692     & 23.791      & -23.285     & -0.54964    & -1.3275     & -11.017     \\ \hline
\textbf{3}    & 69.91       & 29.56       & -12.695     & -0.2148     & 23.954      & -23.056     & -0.52841    & 1.2704      & -10.939     \\ \hline
\textbf{3.25} & 66.122      & 37.914      & -12.796     & -0.93699    & 24.52       & -22.644     & -0.49348    & 2.3896      & -10.565     \\ \hline
\textbf{3.5}  & 63.12       & 44.58       & -12.643     & -1.7067     & 25.404      & -22.115     & -0.45489    & 2.9285      & -9.9496     \\ \hline
\textbf{3.75} & 60.562      & 50.25       & -12.244     & -2.4799     & 26.653      & -21.435     & -0.41539    & 3.2455      & -9.1186     \\ \hline
\textbf{4}    & 58.103      & 55.228      & -11.627     & -3.2211     & 28.243      & -20.589     & -0.37559    & 3.5587      & -8.1204     \\ \hline
\textbf{4.2}  & 56.037      & 58.762      & -11.007     & -3.7601     & 29.774      & -19.798     & -0.34422    & 3.8545      & -7.2472     \\ \hline
\textbf{4.4}  & 53.793      & 61.861      & -10.303     & -4.2231     & 31.558      & -18.921     & -0.31272    & 4.2093      & -6.3499     \\ \hline
\textbf{4.5}  & 52.598      & 63.238      & -9.9291     & -4.4168     & 32.548      & -18.457     & -0.29729    & 4.4153      & -5.9041     \\ \hline
\textbf{4.6}  & 51.351      & 64.488      & -9.545      & -4.5826     & 33.609      & -17.978     & -0.28199    & 4.6426      & -5.4649     \\ \hline
\textbf{4.8}  & 48.704      & 66.573      & -8.7594     & -4.8236     & 35.95       & -16.987     & -0.25223    & 5.1694      & -4.6215     \\ \hline
\textbf{5}    & 45.874      & 68.066      & -7.9703     & -4.9381     & 38.603      & -15.97      & -0.2241     & 5.8018      & -3.8423     \\ \hline
\textbf{5.2}  & 42.902      & 68.915      & -7.1973     & -4.9232     & 41.634      & -14.945     & -0.19782    & 6.5075      & -3.142      \\ \hline
\textbf{5.4}  & 39.839      & 69.133      & -6.4559     & -4.8026     & 44.922      & -13.93      & -0.17367    & 7.3194      & -2.5264     \\ \hline
\textbf{5.5}  & 38.293      & 69.012      & -6.1006     & -4.7111     & 46.674      & -13.429     & -0.16267    & 7.721       & -2.2512     \\ \hline
\textbf{5.75} & 34.42       & 68.017      & -5.2645     & -4.3786     & 51.346      & -12.213     & -0.13688    & 8.5869      & -1.6549     \\ \hline
\textbf{6}    & 30.618      & 66.135      & -4.4838     & -3.9546     & 56.269      & -10.991     & -0.11487    & 8.8965      & -1.1668     \\ \hline
\textbf{6.25} & 26.973      & 63.501      & -3.8169     & -3.4777     & 61.127      & -9.9112     & -0.096243   & 8.1463      & -0.80418    \\ \hline
\textbf{6.5}  & 23.546      & 60.266      & -3.232      & -2.9807     & 65.541      & -8.9079     & -0.080354   & 6.0628      & -0.53395    \\ \hline
\textbf{6.75} & 20.386      & 56.614      & -2.7247     & -2.4879     & 68.805      & -7.9832     & -0.067051   & 2.9286      & -0.33986    \\ \hline
\textbf{7}    & 17.511      & 52.688      & -2.2881     & -2.0149     & 70.557      & -7.1347     & -0.055893   & -0.50182    & -0.20642    \\ \hline
\textbf{7.25} & 14.933      & 48.636      & -1.9148     & -1.5717     & 70.595      & -6.361      & -0.046629   & -3.4089     & -0.12096    \\ \hline
\textbf{7.5}  & 12.652      & 44.574      & -1.5979     & -1.1627     & 69.042      & -5.6568     & -0.038721   & -5.3033     & -0.072013   \\ \hline
\textbf{7.75} & 10.654      & 40.596      & -1.3297     & -0.78932    & 66.251      & -5.0212     & -0.032345   & -6.0878     & -0.050018   \\ \hline
\textbf{8}    & 8.9222      & 36.767      & -1.1046     & -0.45064    & 62.654      & -4.4482     & -0.026907   & -5.9184     & -0.047014   \\ \hline
\textbf{8.5}  & 6.1644      & 29.73       & -0.758      & 0.13364     & 54.491      & -3.4705     & -0.018705   & -3.6832     & -0.075039   \\ \hline
\textbf{9}    & 4.1854      & 23.658      & -0.51707    & 0.61671     & 46.426      & -2.6893     & -0.012911   & -0.23478    & -0.1206     \\ \hline
\textbf{9.5}  & 2.7995      & 18.574      & -0.35141    & 1.0194      & 39.31       & -2.0709     & -0.0089822  & 3.4712      & -0.16427    \\ \hline
\textbf{10}   & 1.8485      & 14.389      & -0.23821    & 1.3607      & 33.173      & -1.586      & -0.0062795  & 6.8979      & -0.19604    \\ \hline
\textbf{11}   & 0.78036     & 8.4159      & -0.10935    & 1.8978      & 23.469      & -0.91716    & -0.0030407  & 11.238      & -0.21617    \\ \hline
\textbf{11.5} & 0.50023     & 6.3549      & -0.074074   & 2.094       & 19.624      & -0.69349    & -0.0021395  & 11.818      & -0.20764    \\ \hline
\textbf{12}   & 0.31812     & 4.7639      & -0.050478   & 2.2339      & 16.321      & -0.52217    & -0.0014307  & 11.627      & -0.19095    \\ \hline
\textbf{12.5} & 0.20114     & 3.5469      & -0.03449    & 2.3102      & 13.495      & -0.39199    & -0.00097355 & 11.059      & -0.16922    \\ \hline
\textbf{13}   & 0.12611     & 2.623       & -0.023693   & 2.322       & 11.093      & -0.29302    & -0.00068698 & 10.371      & -0.14545    \\ \hline
\textbf{13.5} & 0.078825    & 1.9281      & -0.016347   & 2.2656      & 9.1386      & -0.21862    & -0.00047248 & 9.6722      & -0.12169    \\ \hline
\textbf{14}   & 0.04895     & 1.409       & -0.011362   & 2.1616      & 7.4415      & -0.16279    & -0.00031716 & 8.957       & -0.099473   \\ \hline
\textbf{14.5} & 0.030328    & 1.0245      & -0.0079385  & 2.0137      & 6.036       & -0.12097    & -0.00022436 & 8.2266      & -0.079569   \\ \hline
\textbf{15}   & 0.01852     & 0.74138     & -0.005585   & 1.8353      & 4.8794      & -0.090071   & -0.00014818 & 7.4779      & -0.062333   \\ \hline
\textbf{16}   & 0.0071369   & 0.38405     & -0.0027863  & 1.4398      & 3.1612      & -0.050128   & -4.958e-05  & 5.9683      & -0.036691   \\ \hline
\textbf{17}   & 0.0026087   & 0.19726     & -0.001428   & 1.0599      & 2.0256      & -0.028342   & -4.479e-05  & 4.5485      & -0.020357   \\ \hline
\textbf{18}   & 0.00090862  & 0.10066     & -0.00075914 & 0.74191     & 1.2854      & -0.016502   & -3.3907e-05 & 3.3203      & -0.010902   \\ \hline
\textbf{20}   & 0.00010312  & 0.026577    & -0.00018571 & 0.32515     & 0.50633     & -0.0062558  & 1.4992e-06  & 1.5614      & -0.0026851  \\ \hline
\textbf{22}   & 5.8622e-05  & 0.0072509   & -3.5187e-05 & 0.12785     & 0.19932     & -0.0027171  & -1.3178e-05 & 0.61746     & -0.0005809  \\ \hline
\textbf{24}   & 1.8217e-05  & 0.0019795   & -1.5721e-06 & 0.046476    & 5.3331e-06  & -0.0011707  & 0.081897    & 0.20437     & -0.00010039 \\ \hline
\textbf{34}   & -2.8363e-06 & 1.0324e-05  & 1.0305e-05  & 0.00011045  & -1.0479e-05 & -1.4912e-05 & 0.0013272   & 8.8312e-05  & 2.1376e-06  \\ \hline
\textbf{36}   & 9.3668e-06  & 5.0069e-06  & -5.6935e-07 & 2.2347e-05  & 2.9571e-06  & -1.2447e-06 & 0.00051089  & 2.0526e-05  & 5.4374e-06  \\ \hline
\textbf{38}   & 2.606e-06   & 2.613e-06   & -2.1396e-06 & 6.3329e-06  & -2.2547e-09 & -7.0766e-07 & 0.00019677  & 9.2196e-06  & -2.4083e-06 \\ \hline
\textbf{40}   & -1.2331e-06 & 6.182e-07   & 3.8353e-06  & -3.7725e-07 & 1.4089e-06  & 1.8718e-06  & 4.7143e-05  & -1.1627e-06 & 3.9371e-06  \\ \hline
\textbf{44}   & 4.0567e-07  & 4.4677e-08  & -2.7916e-06 & -6.4182e-07 & -9.1605e-08 & 4.8754e-07  & 5.8537e-06  & 8.5205e-07  & -2.4003e-06 \\ \hline
\textbf{46}   & -4.1183e-07 & -9.8637e-07 & -8.8659e-07 & -1.0916e-07 & -6.3187e-08 & 8.6979e-08  & -2.3525e-06 & 3.8519e-06  & -1.8119e-07 \\ \hline
\textbf{48}   & -6.5671e-08 & -1.7879e-07 & -3.77e-08   & -1.2253e-08 & -9.7311e-09 & 1.4769e-08  & 4.3164e-06  & -1.3175e-07 & -2.3639e-08 \\ \hline
\textbf{50}   & -1.066e-08  & -3.3135e-08 & -3.466e-09  & -4.2342e-11 & -2.5201e-09 & 3.2097e-09  & -1.6712e-08 & -2.0993e-08 & -4.4837e-09 \\ \hline
\end{tabular}%
}
\caption{Tabular values of the hyperfine coupling constant for the 13 lowest lying states of the K-He complex. Values of R (in bohr) and $\alpha$ (in MHz).}
\label{tab:alpha_table}
\end{table*}
In table~\ref{tab:alpha_table} we list the calculated values of the hyperfine-coupling constant $\alpha\left(R\right)$ of the first 13 lowest lying states of the K-He complex, as shown in Fig.~\ref{fig:all_rhos}.

\clearpage

\section*{S3. Validation of Calculation Results}
\subsection*{Comparison with Other Models}

\begin{figure*}[ht!]
\centering
   
        \includegraphics[width=0.7\textwidth]{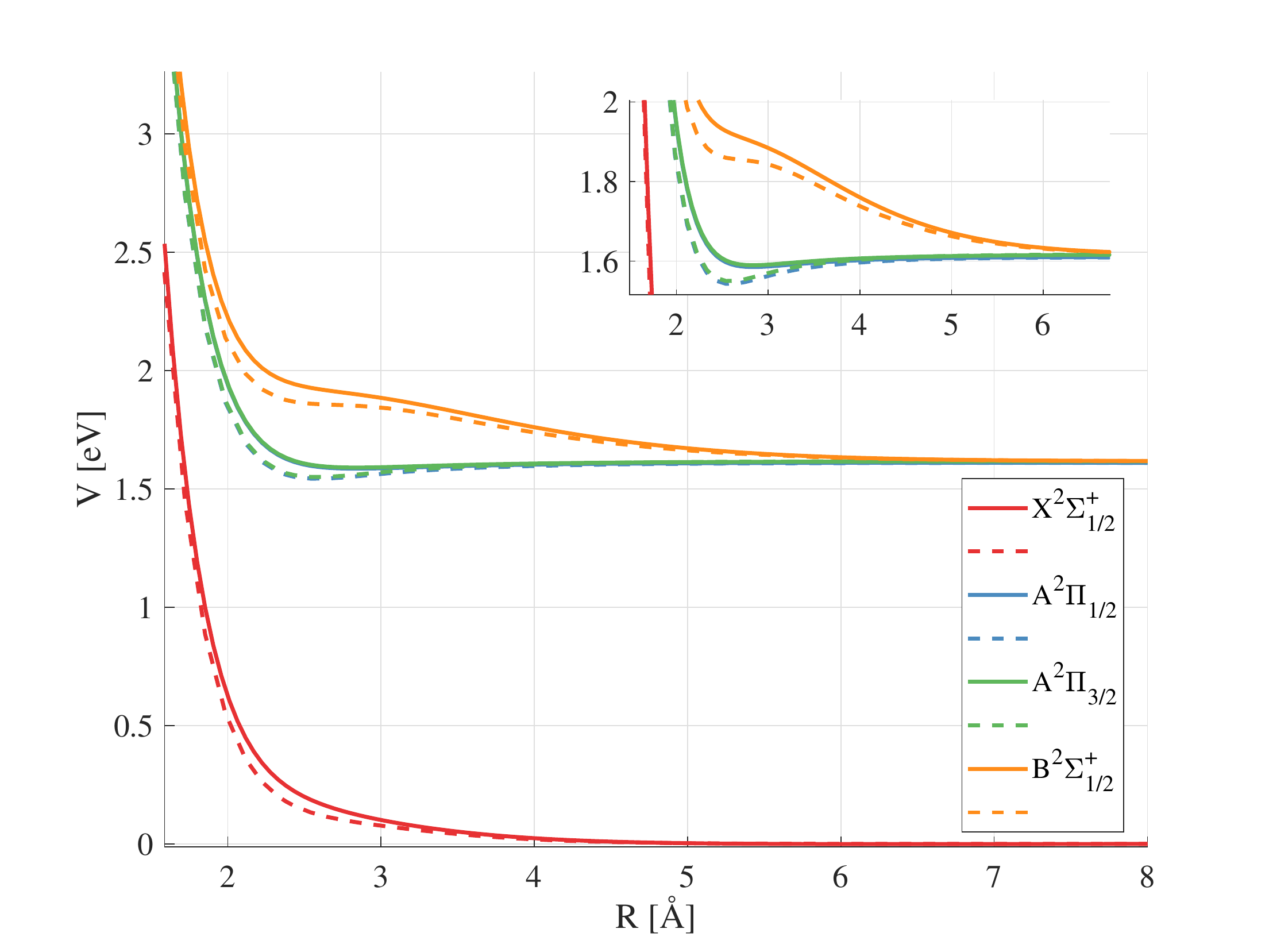}
\caption{The PECs for the ground state and two lowest lying excited states of the K-$^3$He complex. Each curve is labeled by its molecular term symbol and its corresponding asymptotic K atom configuration. The solid lines are the results of Blank et.~al.~\cite{blank_comparison} whereas the dashed lines are the results of this work. The inset shows that significant discrepancies ($\sim$10 meV) only occur at distances smaller than about 3 $\text{\AA}$, that have little bearing on the spin exchange rate for typical collision energies.
}
\label{fig:PECs_vs_blank}
\end{figure*}

We validated our calculations of the potential energy curves (PECs) for the K-$^3$He complex by comparison with other references. In Figure~\ref{fig:PECs_vs_blank} we present the calculated PECs of the ground-state as well as the PECs associated with the 4p states of the K atom with the PEC in Ref.~\cite{blank_comparison}. The various PEC are reproduced reasonably well at distance higher than 3 $\text{\AA}$. The discrepancy between the calculations is limited to below 3 $\text{\AA}$, whose effect on the spin dynamics at collisions is negligible.

\begin{figure*}[ht!]
\centering
        \includegraphics[width=0.7\textwidth]{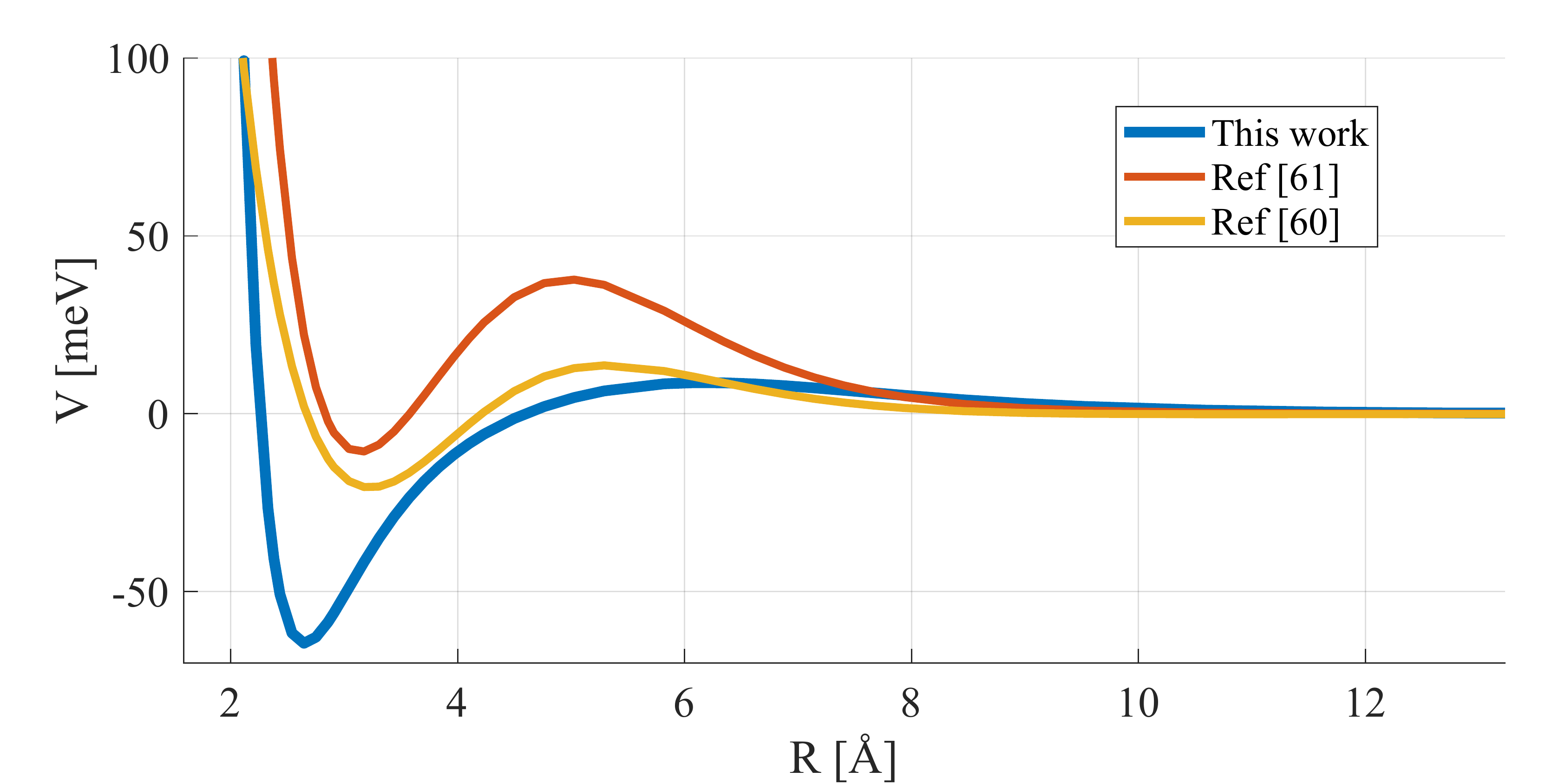}
\caption{The PECs for the excited state associated with the 5s state of the K atom from this work and from Refs.~\cite{chattopadhyay2012comparative,kontar2017electronic.}
}
\label{fig:5s_PEC_comparison}
\end{figure*}
In Figure~\ref{fig:5s_PEC_comparison} we present the PEC associated with the 5s state of the K atom calculated in this work (blue) and compare it with the relevant PEC from Ref.~\cite{chattopadhyay2012comparative} (yellow) and Ref.~\cite{kontar2017electronic} (red).

The potentials show qualitative agreement, with all three estimates exhibiting the potential well and barrier, with comparable well depths: $35\;\mathrm{meV}$ for Ref.~\cite{chattopadhyay2012comparative}, $46\;\mathrm{meV}$ for Ref.~\cite{kontar2017electronic} and $76\;\mathrm{meV}$ in this work. As mentioned in the main text, despite the quantitative differences between the potentials, the resonance-based mechanism of SEOP is robust; calculation of the thermally-averaged polarization rate-coefficient $k_{\mathrm{se}}^\mathrm{(q)}$ at $100\,\mathrm{^\circ C}$ for $\gamma^{-1}=1\;\mathrm{ns}$ produces sizeable rate coefficients for scattering of this excited state: $k_{\mathrm{se}}^\mathrm{(q)}=615\times10^{-20}\mathrm{cm}^3/\mathrm{s}$ for this work, $k_{\mathrm{se}}^\mathrm{(q)}=543\times10^{-20}\mathrm{cm}^3/\mathrm{s}$ for Ref.~\cite{kontar2017electronic} and $k_{\mathrm{se}}^\mathrm{(q)}=305\times10^{-20}\mathrm{cm}^3/\mathrm{s}$ for Ref.~\cite{chattopadhyay2012comparative}.

The quantitative discrepancies in the exact shapes of the potentials possibly originate from differences in the calculations:
\begin{enumerate}
  \item Both Refs.~\cite{chattopadhyay2012comparative,kontar2017electronic} employ an open-shelled HF reference function, which suffers from spin-contamination, whereas in this work we use EOM-CCSD which uses the closed-shell, and therefore spin-pure, $\mathrm{\left(K-He\right)^+}$ cation reference function.
  \item The basis set used for the potassium atom in Ref.~\cite{kontar2017electronic} (def2-SVPD) is smaller than that used in this work (aug-pc-3.)
  \item The basis set used in this work is also significantly more contracted than the basis sets used in both Ref.~\cite{kontar2017electronic} (def2-SVPD) and in Ref.~\cite{chattopadhyay2012comparative} (def2-QZVPPD), being composed of many more primitive gaussians, able to better capture features of the wavefunction close to the atom core.
\end{enumerate}

It is also interesting to compare the asymptotic energy of the PEC associated with the 5s state of the K atom to the experimental value. Our work agrees within $\mathrm{0.6\%}$, Ref.~\cite{kontar2017electronic} agrees to $\mathrm{1.3\%}$ but the value in Ref.~\cite{chattopadhyay2012comparative} is not reported in a way that facilitates comparison.

\clearpage
\subsection*{Comparison with Experiment}

To validate the calculation of the hyperfine coupling constant, we compared our calculated values with several references. First, we verified the basic properties such as the orbital symmetries, that the integral over all space gives $n$ electrons for the charge density, and a single electron for the spin density.
Second, we estimated the frequency-shift enhancement factor for the ground state $\kappa_{0,\mathrm{theory}}=6.64\; (200\;\mathrm{^\circ C})$ (using the spin density of the ground state, $\rho_\mathrm{spin}^{4S\;:\;^2\Sigma}$ as proposed in Ref.~\cite{schaefer1989frequency}) and compared it with the experimental result $\kappa_{0,\mathrm{exp}}=6.01\pm0.01\; (200\;\mathrm{^\circ C})$ measured in Ref.~\cite{babcock2005he}. This new estimate improves on the previous theoretical estimate of $\kappa_{0,\mathrm{theory}}=8.5$ in Ref.~\cite{walker1989estimatesPRA}.
Third, we estimated the spin exchange rate coefficient for the ground-state, using the semi-classical approximation, yielding $\tilde{k}_{\mathrm{SE}} = 8.2\times10^{-20}\frac{\text{cm}^3}{\text{sec}} $ at $200 ~\mathrm{^\circ C}$. This result is in a reasonable agreement with the following experimental results $k_{\mathrm{se}} = 6.1(4)\times10^{-20}\frac{\text{cm}^3}{\mathrm{sec}} $ in Ref.~\cite{babcock2003hybrid}, $k_{\mathrm{se}} = 5.5(2)\times10^{-20}\frac{\text{cm}^3}{\mathrm{sec}} $ in Ref.~\cite{babcock2005spin},  $k_{\mathrm{se}} = 6.1(7)\times10^{-20}\frac{\mathrm{cm}^3}{\mathrm{sec}} $ in Ref.~\cite{walker2010method}, and $k_{\mathrm{se}} = 7.5(5)\times10^{-20}\frac{\mathrm{cm}^3}{\mathrm{sec}} $ in Ref.~\cite{singh2015development}.

\clearpage

\end{document}